\documentclass[showpacs,preprintnumbers,amsmath,amssymb,aps, prd]{revtex4}
\usepackage{graphicx}% Include figure files
\usepackage{grffile}
\usepackage{mathrsfs, mathtools}
\usepackage{caption}
\usepackage{float}
\usepackage{xcolor}
\usepackage{changepage}
\usepackage{dcolumn}% Align table columns on decimal point
\usepackage{bm}% bold math
\usepackage{graphicx,subfigure,epsfig}
\usepackage{multirow}
\usepackage{ifpdf}
\begin{document}
\title{An atom in front of  Lorentz violating Kalb-Ramond black hole background}
\author{Anisur Rahaman}
\email{manisurn@gmail.com} \affiliation{Durgapur Government
College, Durgapur-713214,  India}
\date{\today}% It is always \today, today,
\begin{abstract}
 We investigate the role of Lorentz violation in the acceleration radiation
 produced when an atom falls into a Kalb-Ramond (KR) black hole and observe
 that the amplitude and an exponential (Planck-like) factor are both shaped
  by the Lorentz-violating parameter, indicating a breach of the equivalence
  principle and resembling characteristics observed in bumblebee gravity models.
   We further investigate how Lorentz violation and conformal symmetry work
    together to determine the thermodynamic behavior of the system and the
    implications for the equivalence principle by looking at the transition
     probabilities of a two-level atomic detector interacting with the black hole.
     These findings provide new information about the interaction of black hole
      entropy, symmetry breaking, and possible observational probes of novel physics
      beyond general relativity. The horizon brightening acceleration radiation (HBAR) entropy in the KR black hole spacetime is also calculated in detail. Although the corrections are very different from those in bumblebee gravity, our study demonstrates that even though Lorentz-violating events alter the entropy, it nevertheless maintains a structural resemblance to the ordinary Bekenstein-Hawking entropy.
\end{abstract}
%\verb+\pacs{#1}+ command.\end{abstract}
%\pacs{11.10.Ef, 11.30.Rd}
\maketitle
 \section{Introduction}
Black holes are thought of as accurate solutions to  highly non-linear
field equations of the fundamental theory  of general relativity, which were
initially put out by Einstein \cite{EIN15, EIN16}. These principles offer the
framework that integrates geometry and gravity. The amazing combination of general
relativity and thermodynamics produced a number of ground-breaking findings fifty
years after Einstein's original theory. Black hole thermodynamics
\cite{HAW1, HAW2, BAKE1, BAKE2}, Hawking radiation \cite{HAW1, HAW2, BAKE1},
 and the more general study of particle emission from black holes \cite{PAGE1, PAGE2, PAGE3}
 were all significantly impacted by Stephen Hawking's seminal works \cite{HAW1, HAW2}, which showed that black holes radiate when quantum effects in curved spacetime are taken into account.In addition to these, the Unruh effect \cite{UNRH} and the acceleration radiation phenomenon \cite{DEWITT, UNRUH2, MULLER, Vanzella, Higuichi, ONRDENZ1, ONRDENZ2, ONRDENZ3, ONRDENZ4} have become essential elements in comprehending quantum fields close to powerful gravitational sources.

One particularly fascinating aspect of this domain is the study of geometrical
and thermal characteristics of the horizons of black hole and how closely they
relate to the dynamics of particles and detectors, like two-level atomic system,
 in these harsh settings. The Unruh effect \cite{UNRH}, which is the observation of
  particles by an accelerating observer in an inertial vacuum, is closely related to
  Hawking's prediction of thermal radiation. Recent studies have demonstrated that,
   largely independent of the particular black hole metric, atoms falling into black
    holes exhibit thermal radiation effects similar to Hawking radiation
    \cite{PAGE0, WEISS, PHILBIN}.  A A Few recent work in this line are reported
    in the papers \cite{RITU, BRM1, BRM2} A useful framework for tying together general relativity,
    quantum field theory, and atom optics has also been made available by the advent
    of ideas like Horizon Brightened Acceleration Radiation (HBAR) Entropy \cite{PAGE0}.

he analysis of near-horizon black hole dynamics has also focused a lot of attention
on conformal symmetry \cite{CONF1, CONF2, ONRDENZ1, ONRDENZ2, ONRDENZ3, ONRDENZ4},
 exposing rich symmetry structures that influence both classical and quantum aspects
 of black hole spacetimes. However, the integration of general relativity with
  quantum mechanics continues to be one of the most difficult unresolved issues
  in contemporary physics. No comprehensive answer has been found despite a number
  of theoretical attempts, ranging from string theory to canonical quantum gravity,
  to find a consistent theory \cite{ROVELLI, CARLIP, ACV, KOPAPR}.

 Kalb-Ramond (KR) gravity \cite{KALBRAM1, KALBRAM2, KALBRAM3, KRreVIEWS} provides
 a particularly strong alternative paradigm in this regard. The Kalb-Ramond field,
 which has its roots in string theory, offers a strong mechanism for spontaneous
 Lorentz symmetry violation by introducing an antisymmetric rank-2 tensor field
  that relates to spacetime geometry naturally. The gravitational landscape is reshaped,
  and geodesic structures are modified by this nontrivial background
  field \cite{KR, KRLOreNTZ1, KRLOreNTZ2, KRLOreNTZ3}. The substantial
  impacts of the KR field on black hole spacetimes have been studied recently;
   in particular, how it alters the thermodynamics, photon dynamics,
   and structure of Reissner-NordstrÃ¶m-like black holes. For example,
   the black hole horizon, curvature invariants, Hawking temperature,
    and entropy are all impacted by the Lorentz-violating parameter that
    the KR field introduces. Research has also looked at how the KR field
    affects photon propagation in both vacuum and plasma, identifying
    changed gravitational lensing signatures and refractive effects \cite{PAGE0}.

 Another method for incorporating Lorentz violation while taking the
 indirect quantum effect into account is bumblebee gravity \cite{BB0, BB1, BB20}.
 The effect of Lorentz violation on several physical processes has
 also been studied in the modern era
  \cite{BB2,BB3, BB4, BB5, BB6, BUMBLES1,BUMBLES2, BUMBLES3, BUMBLEBEE}.
   Additional observational research has limited the KR framework through
   the analysis of optical effects and geodesic precession, contrasting
   them with rival Lorentz-violating models such as the Bumblebee (BM)
   field \cite{KRBLACKHOLE1, KRBLACKHOLE2, KRBLACKHOLE3}. Observable
   fingerprints that could potentially differentiate these unusual
   circumstances from conventional black holes are suggested by variations in
    quasinormal modes, ring-down wave forms, Hawking radiation, and gravitational
    lensing between KR and other black holes.

A particularly fascinating and still underexplored frontier
concerns the acceleration radiation experienced by an atom falling
into a black hole within the Kalb-Ramond framework. Although
acceleration radiation has been studied extensively across a
variety of black hole metrics, incorporating quantum corrections
and accounting for the role of antisymmetric tensor fields remain
key challenges \cite{KRBLACKHOLE1, KRBLACKHOLE2, KRBLACKHOLE3}.
Notably, the mechanism of spontaneous breaking of Lorentz symmetry
inherent to KR gravity introduces subtle but potentially profound
modifications to the radiation processes encountered by test
particles and detectors near black holes. This motivates a
detailed investigation of how an atom  acting effectively as an
Unruh-DeWitt detector perceives acceleration radiation in the
presence of a Kalb-Ramond black hole.

Furthermore, the ramifications go far into fundamental physics. A
fundamental tenet of general relativity, the equivalence principle
states that inertial and gravitational effects are identical
locally Nevertheless, studies have demonstrated that explicit
breaches of this principle across many spacetime geometries and
vacuum states can be found using the response function of an
Unruh-DeWitt detector \cite{SINGLETON, SINGLETON1, VEQUIV}. This
brings up a crucial and important query: what is the impact of
Lorentz symmetry violation in Kalb-Ramond gravity on the futures
of equivalence principle? Investigating this junction should
provide information about the basic symmetries that underlie
spacetime itself as well as the nature of acceleration radiation.

Collectively, these investigations provide a complex pictureof the
relationship between modified gravity, quantum field theory, and
general relativity, offering a wealth of opportunities for
theoretical investigation as well as possible experimental
validation. Atom acceleration research close to Kalb-Ramond black
holes provides a fresh and illuminating insight into these two
phenomena, with the potential to significantly improve our
comprehension of both gravity and quantum theory.

\section{A brief description of Lorentz violating Kalb-Ramond spacetime background}
 A static and spherically symmetric solutions for Schwarzschild-like configurations
 with and without implementing the cosmological constant, were obtained in \cite{KR},
 taking into account the KR field. The initial version of KR Gravity was created
 in \cite{KALBRAM1}. The KR field in these solutions obtains a non-zero vacuum
 expectation value, signifying the gravity's spontaneous Lorentz symmetry violation.
 In the Einstein-Hilbert action, their approach incorporates a self-interacting
  KR field that is not minimally linked \cite{KRKOS, KRM}.  The action comes up
  with the mathematical expression

\begin{eqnarray}
 \mathcal{A} =\int d^4x\sqrt{-g}\Bigg[ R-\frac{1}{6}H^{\mu\nu\rho}H_{\mu\nu\rho}
 -V(B^{\mu\nu}B_{\mu\nu})\nonumber\\
 +g_0 B^{\rho\mu}B^{\nu}_{\mu}R_{\rho\nu}+g_1 B^{\mu\nu}B_{\mu\nu}R\Big].
\label{ACT0}
\end{eqnarray}
In Eqn. (\ref{ACT}),  $g_0$ and $g_1$ are two coupling constant, $B_{\mu\nu}$ denotes the
 KR field, and the
field strength tensor of the KR field   $B_{\mu\nu}$ is
\begin{equation}
H_{\mu\nu\rho}=\partial_{[\mu}B_{\nu\rho]}= \partial_\mu B_{\nu\rho} + \partial_\nu B_{\mu\rho} + \partial_\rho B_{\mu\nu}
\end{equation}
The self-interaction potential $V(B^{\mu\nu}B_{\mu\nu})$ that  depends
on $B^{\mu\nu}B_{\mu\nu}$ has been introduced in order to maintain the invariance of the theory
upon observer local Lorentz transformations. In order to generate a non-vanishing vacuum
expectation value (VEV) for the KR field, $\langle B_{\mu\nu} \rangle= b_{\mu\nu}$, a potential with a general
form  $V = V({\mu\nu} \mp b^2)$ is chosen. Consequently, VEV  is determined by the
 condition $b_{\mu\nu}b^{\mu\nu} \mp b^2=0$. Given this vacuum expectation value of
 the KR field  the coupling term containing $g_0$
in the action (\ref{ACT0}) reduces to a linear term in $R$, which
can then be transformed into an Einstein-Hilbert term through a
coordinate transformation. Variation of the action with respect to the
metric now yields
\begin{eqnarray}
&& R_{\mu\nu}-\frac{1}{2}g_{\mu\nu}R=\frac{1}{2}H_{\mu\alpha\beta}H_{\nu}^{\;\;\alpha\beta}-\frac{1}{12}g_{\mu\nu}H^{\alpha\beta\rho}H_{\alpha\beta\rho}\nonumber\\
&& +2V'(X)B_{\alpha\mu}B^{\alpha}_{\;\;\nu}-g_{\mu\nu}V(X)+g_0\Bigg[\frac{1}{2}g_{\mu\nu}B^{\alpha\rho}B^{\beta}_{\;\;\rho}R_{\alpha\beta}\nonumber\\
&& -B^{\alpha}_{\;\;\mu}B^{\beta}_{\;\;\nu}R_{\alpha\beta}-B^{\alpha\beta}B_{\nu\beta}R_{\mu\alpha}-B^{\alpha\beta}B_{\mu\beta}R_{\nu\alpha}\nonumber\\
&&  +\frac{1}{2}\nabla_{\alpha}\nabla_{\mu}\left(B^{\alpha\beta}B_{\nu\beta}\right)+\frac{1}{2}\nabla_{\alpha}\nabla_{\nu}\left(B^{\alpha\beta}B_{\mu\beta}\right)\nonumber\\
&&
-\frac{1}{2}\nabla^{\alpha}\nabla_{\alpha}\left(B_{\mu}^{\;\;\rho}B_{\nu\rho}\right)-\frac{1}{2}g_{\mu\nu}\nabla_{\alpha}\nabla_{\beta}\left(B^{\alpha\rho}B^{\beta}_{\;\;\rho}\right)\Bigg].
\end{eqnarray}
We are familiar with the  spherical  symmetric  metric in a general form
\begin{eqnarray}
ds^2=-A(r)dt^2+B(r)dr^2+r^2\left(d\theta^2+\sin^2\theta
d\phi^2\right),
\end{eqnarray}
where $A(r)$ and $B(r)$ are lapse functions which have crucial link with nature of
space time curvature. It is convenient to define a  2-form $B_{2}=\frac{1}{2}B_{\mu\nu}(r)dx^\mu\wedge dx^\nu$.  Now if we
choose $V(x)=\lambda x^2/2$, where $x=B^{\mu\nu}B_{\mu\nu}+b^2$,
we find $V'\equiv 0$. Ultimately we obtain  $b^{\mu\nu}b_{\mu\nu}=-b^2$.
Therefore, the equations of motion that follows from the action
(\ref{ACT}) are provided by
\begin{eqnarray}
2\frac{A''(r)}{A(r)}
-\frac{A'(r)B'(r)}{A(r)B(r)}-\frac{A'(r)^2}{A(r)^2}
+\frac{4A'(r}{rA(r)}=0,\label{EQM1}
\end{eqnarray}
\begin{eqnarray}
2\frac{A''(r)}{A(r)}
-\frac{A'(r)B'(r)}{A(r)B(r)}-\frac{A'(r)^2}{A^2}
-\frac{4B'(r)}{rB(r)}=0,\label{EQM2}
\end{eqnarray}
\begin{eqnarray}
2\frac{A''(r)}{A(r)}
-\frac{A'(r)B(r)'}{A(r)B(r)}-\frac{A'(r)^2}{A(r)^2}
+\frac{(1+l)}{lr}
\left(\frac{A'(r)}{A(r)}-\frac{B'(r)}{B(r)}\right)
 -2\frac{B(r)}{lr^2}+\frac{2(1-l)}{lr^2}=0. \label{EQM3}
\end{eqnarray}
where $l$ is defined by  $l=g_1 b^2/2$. Now subtracting Eqn.
(\ref{EQM1}) from Eqn.(\ref{EQM2}) we have
\begin{eqnarray}
\frac{A'(r)}{A(r)}+\frac{B'(r)}{B(r)}=0 \label{SUB}
\end{eqnarray}
The Eqn. (\ref{SUB}) renders a  solution by  which we can express
$B(r$) in terms of $A(r)$ or viceversa.
\begin{eqnarray}
 B(r)=\frac{\tilde{C}}{A(r)}. \label{AINB}
\end{eqnarray}
If we plug in Eqn. (\ref{AINB})  into Eqn. (\ref{EQM3}), it solves
$A$ after a few steps of algebra and we land on
\begin{eqnarray}
A(r)=\frac{\tilde{C}}{1-l}+\frac{C}{r}.
\end{eqnarray}
The constant $C$ has been evaluated using the Komar mass, which results in $C=-\frac{2M}{c^2}$.
We require $\tilde{C}=1$ in order to agree with the Schwarzschild solution.
 Naturally, the coupling with the KR field must be set to zero, or $l=0$.
 In this case, we finally arrive at the following solution for the specified action (\ref{ACT}).
\begin{eqnarray}
ds^2 =
-\left(\frac{1}{1-l}-\frac{2M}{rc^2}\right)dt^2+\left(\frac{1}{1-l}-\frac{2M}{rc^2}\right)^{-1}dr^2
+ r^2 d\Omega^2.\label{KR0}
\end{eqnarray}
 where
\begin{eqnarray}
d\Omega^2=d\theta^2+\sin^2\theta d\phi^2.
\end{eqnarray}
In the significant article \cite{KR}, this solution was initially reported.
The additional parameter $l$ in this case describes the spontaneous violation of
 Lorentz symmetry.  Notably, the new parameter $l=g_0 b^2/2$, which is the product
  of the coupling constant and the square of the KR field's expectation value,
   may potentially be predicted to have any reasonable value. It is necessary
   to adhere to the restriction $0\leq l< 1$ in order to safeguard the signature
   of the metric $(-,+,+,+)$ within the event horizon. Asymptotically,
   the answer is flat.  It will be more noticeable as $l \to 1$.
   The solution deviates from the Minkowski solution at infinity.
\begin{eqnarray}
A(r)_{r=r_e}= 0 \Rightarrow  \frac{1}{1-l}+\frac{2GM}{rc^2}=0.
\label{EHOR}
\end{eqnarray}
which yields
\begin{eqnarray}
r_{e}= \frac{2GM (1 - l)}{c^2} \label{HOR}
\end{eqnarray}
The Kretschmann scalars for the KR BH ($K_{KR}$) is given by
\begin{equation}
\mathcal{K}_{KR}=\frac{48 M^2}{c^4r^6}-\frac{16 l M}{(1-l)c^2 r^5}+\frac{4
l^2}{(1-l)^2 r^4}.
\end{equation}
The Hawking temperature is given by
\begin{eqnarray}
\mathcal{T}_H &=& \frac{c\hbar}{4\pi k_B }A'(r_{e})\nonumber \\
&=&\frac{\hbar c^3}{8\pi k_{B}GM(1-l)^2}
\end{eqnarray}
\section{The KR black hole in Rindler configuration}
Let us begin by considering the Minkowski metric in a
(1+1)-dimensional spacetime, whose line element is given by
\begin{equation}
ds^2=c^2dt^2-dx^2. \label{MIN}
\end{equation}
If a body moves with uniform proper acceleration `$a$' in this
flat (1+1)-dimensional spacetime, its position and time
coordinates, as functions of the proper time $\tau$, can be
provided by
\begin{align}
x(\tau)&=\frac{c^2}{a}\cosh(\frac{a\tau}{c}),\label{T1}
\end{align}
\begin{align}
t(\tau)&=\frac{c}{a}\sinh(\frac{a\tau}{c}).\label{T2}
\end{align}
These expressions (\ref{T1} , \ref{T2}) satisfy the Minkowski
metric equation (\ref{MIN}) by construction. Now, let us introduce
the following coordinate transformations:
\begin{align}
x&=\eta\cosh(\frac{\mathbf{a}\mathbf{t}}{c}),\label{TT1}
\end{align}
\begin{align}
t&=\frac{\eta}{c}\sinh(\frac{\mathbf{a}\mathbf{t}}{c}).\label{TT2}\\
\end{align}
Under the transformation  (\ref{TT1}) and (\ref{TT2}) the metric
(\ref{MIN}) transforms into the Rindler form:
\begin{equation}
ds^2=\left(\frac{\mathbf{a}\eta}{c^2}\right)^2c^2d\mathbf{t}^2-d\eta^2.\label{RMET}
\end{equation}
This Rindler metric (\ref{RMET}) describes the view point of a
uniformly accelerating observer. By comparing (\ref{T1}) with
(\ref{TT1}), we find that the proper time $\tau$ can be related to
the new coordinate time $x$ and $t$
  as:
\begin{equation}
\tau=\left(\frac{\mathbf{a}}{a}\right)\mathbf{t}\label{PRT}
\end{equation}
Furthermore, comparing equations (\ref{T2}) and (\ref{TT2}), we
obtain the expression for the uniform acceleration in Minkowski
spacetime:
\begin{equation}
a=\frac{c^2}{\eta}. \label{UNIAC}
\end{equation}
We now try to reconstruct the Lorentz-violating KR black hole
metric close to the event horizon in the Rindler form using this
mathematical framework. This will enable us to derive the formula
for the uniform acceleration that a body falling freely near the
event horizon experiences. The lapse function $A(r)=0$ (as
mentioned in Eqn.(\ref{HOR})) can be set to find the event horizon
of the Lorentz-violating KR black hole. Notably, because it
depends on the parameter $l$, the location of the event horizon in
this spacetime is different from that of the typical Schwarzschild
black hole.

In order to continue, we expand the metric in the near-horizon.
In particular, we increase $A(r)$ as a Taylor series about the event
 horizon radius $r_{e}$ while keeping terms in the tiny parameter
 $(r-r_{e})$ up to first order. It provides

\begin{equation}
A(r)\cong
A(r_{r_{e}})+(r-r_{e})\frac{dA(r)}{dr}\biggr|_{r=r_{e}}=(r-r_{e})A'(r_{e}).
\label{EXPHOR}
\end{equation}
Using this, we write the (1+1)-dimensional line element near the
horizon as
\begin{equation}
\begin{split}
ds^2&=A(r)c^2dt^2-A(r)^{-1}dr^2\\
&\cong(r-r_{e})A'(r_{e})c^2dt^2-\frac{1}{(r-r_{e})A'(r_{e})}dr^2.
\label{APPDS}
\end{split}
\end{equation}
To express (\ref{APPDS}) in the Rindler form, we introduce a new
coordinate $\eta$ which has the definition
\begin{equation}
\eta=2\sqrt{\frac{(r-r_{e}))}{A'(r_{e})}}.\label{ETA}
\end{equation}
Substitution of  the expression of $\eta$ into the metric (\ref{APPDS})
yields
\begin{equation}
ds^2\cong \frac{\eta^2A'^2(r_{e})}{4}c^2dt^2-d\eta^2,
\label{RINDSCH}
\end{equation}
which manifests in the Rindler form on the metric in $(1+1)$  dimension.
Here, the derivative  $A'(r_{e})$ is given by
\begin{equation}
A'(r_e) = \frac{c^2}{2 M G (1 - l)^2},
\end{equation}
since the event horizon radius $r_{e}$ is provided by
\begin{equation}
r_{e} = \frac{2MG(1 - l)}{c^2}. \label{ORF}
\end{equation}
By comparing equation (\ref{RINDSCH}) with the general Rindler
metric (\ref{RMET}), we identify the uniform acceleration
associated with curves of constant $\eta$ as:
\begin{equation}
\begin{split}
a&=\frac{c^2}{\eta}=\frac{c^2}{2\sqrt{\frac{r-r_{e}}{A'(r_{e})}}}\\
&\cong\frac{c^2\sqrt{A'(r_{e})}}{2\sqrt{r}}\left(1+\frac{r_{e}}{2r}\right)
\end{split}.\label{DEFAC}
\end{equation}
Keep in mind that we have only kept the first order term of Taylor's series.
 This last solution gives the uniform acceleration, stated in terms of the
 near-horizon parameters, that a body near the horizon of the
 Lorentz-violating KR black hole would experience.

\section{Solution of Klein-Gordon field equation in the Lorentz violating KR background}
We consider the spherically symmetric bumblebee gravity inspired
Lorentz violating geometry in four spacetime dimensions. The
metric of which  is given by
\begin{equation}
ds^2=-A(r)c^2dt^2+A(r)^{-1}dr^2+r^2d\Omega^2. \label{LVSCHD}
\end{equation}
Here, $A(r)= \frac{1}{1-l}-\frac{2MG}{rc^2}$. If we  take into
account the minimal coupling of this background  with scalar field
$\Phi$ the action in that case reads
\begin{equation}
{\cal A}= -\frac{1}{2}\int d^4 x
\sqrt{-g}[g^{\mu\nu}\partial_\mu\Phi\partial_\nu\Phi+ m_0^2\Phi].
\label{ACT}
\end{equation}
The  Klein-Gordon type  equation for the scalar field that follows
from Eqn. (\ref{ACT}) for photon $(m_0=0)$ is
\begin{equation}
\frac{1}{\sqrt{-g}}\partial_\mu(\sqrt{-g}g^{\mu\nu}\partial_\nu\Phi)\Phi=0.
\label{KGE}
\end{equation}
For photon the  equation with temporal  and special part is given
by
\begin{equation}
\frac{1}{T(t)}\frac{d^2T(t)}{dt^2}-\frac{A(r)}{r^2R(r)}
\frac{d}{dr}\left(r^2A(r)\frac{dR(r)}{dr}\right)=0, \label{KGST}
\end{equation}
where the general form wave function $\Phi(t,r)$ of the photon can
be written down as
\begin{equation}
\Phi(t,r)=T(t)R(r),
\end{equation}
bearing in mind that, in this instance, the spatial and temporal components of
the solution are separable.
The general solution of Eqn.(\ref{KGST}) can be written down as
\begin{equation}
\Phi_\nu(t,r)=\exp\left[- i\nu t+ i\nu\int\frac{dr}{A(r)}\right].
\label{SOLN}
\end{equation}
The frequency of the emitted photon near the Lorentz violating KR black hole is described by
$\nu$ in Eqn.(\ref{SOLN}). In this study, we adopt the model presented in \cite{FULLING},
which uses a mirror to protect Hawking radiation emitted from the black hole. This configuration
is the same as a Boulware vacuum \cite{BOUL}. Now  an operator
$\hat{\zeta}=|g\rangle\langle e|$ is defined, and hence, of the atom field interaction
term associated to the
the Hamiltonian  the  can be written down as
\begin{equation}
\hat{H}_I(\tau)=\hbar \mathcal{G}[\hat{a}_\nu
\Phi_\nu(t(\tau),r(\tau))+H.c.][\hat{\zeta}e^{-i\Omega\tau}+h.c.].
\label{INT}
\end{equation}
The atom-field coupling constant  $\mathcal{G}$, the  annihilation operator
associated with to photon  $\hat{a}_\nu$ having frequency $\nu$ represented by
the wave function $\Phi_\nu$, and obviously $\hat{a}_\nu^\dagger$ is the
corresponding creation operator, and the atomic frequency  $\Omega$ describe
the atom field interaction Hamiltonian in Eqn.(\ref{INT}).
When a scalar photon is simultaneously released, the probability of excitation
of the associated atom is determined by
\begin{equation}
\begin{split}
P_{e} &= G^2\left|\int_{r_{e}}^{\infty} dr \frac{dr}{d\tau}
e^{i\nu t(r) - i\nu \tilde r(r)}
e^{i \Omega \tau(r)}\right|^2 \\
&=\left| G^2 \int_{r_+}^{\infty} dr\, \frac{1}{\sqrt{1 - A(r)}}
 e^{-i\nu \int \frac{dr}{A(r)\sqrt{1 - A(r)}} - i\nu \int
 \frac{dr}{A(r)}}\right|^2
\end{split}
\label{PE}
\end{equation}
We take into consideration a generic example in which the event horizon
connected to the lapse function
$A(r)$ with black hole geometry is $r_{e}$. The previously defined expressions
of $\tau(r), t(r)$
are employed in equation (\ref{PE}) to produce the explicit form of excitation the
probability and $\tilde{r}$ reads
\begin{equation}
\tilde{r}= \int \frac{dr}{A(r)}
\end{equation}
With regard to the Lorentz violating spherically symmetric KR metric in the Rindler form,
 we can now implement
a near-horizon expansion.  In this case, the Taylor series expansion of $A(r)$
about the event horizon $r_{e}$,
which is provided in Eqn. (\ref{EXPHOR}), will be used. Note that the terms up to
the first order in the near-horizon
expansion parameter $(r-r_{e})$ will be utilized when the near-horizon expansion is executed.
 It should be noted that Eqn. (\ref{APPDS}) already provides the metric in $(1+1)$-dimensional spacetime
within the Rindler form.
and have previously defined a new coordinate $\eta$, which is provided in
 Eqn. (\ref{ETA}), in order
to express (\ref{APPDS}) into the Rindler form. Thus, the
transition probability using the near-horizon approximation reads
\begin{equation}
P_{e} = G^2 \left|\int_{r_{e}}^{\infty} dr \frac{e^{-i\nu \int
\frac{dr}{(r - r_{e}) r(r_{e})}}}{ \sqrt{1 - (r - r_{e})
A'(r_{e})}}{\sqrt{1 - (r - r_{e}) A'(r_{e})}} \times e^{-i\nu \int
\frac{dr}{(r - r_{e}) A'(r_{e})}} \, e^{-i\Omega \int
\frac{dr}{\sqrt{1 - (r - r_{e}) A'(r_{e})}} }\right|^2.
\end{equation}
Let us introduce a new variable $k$ which is defined by
\begin{equation}
\Omega(r-r_{e)}= k \label{NV}
\end{equation}
and assume that $\Omega >> A'(r_{e}), \nu$ and $k$. If we implement
 Eqn. (\ref{NV}) in Eqn. (\ref{PE}) the elicitation probability
turns into
\begin{eqnarray}
P_{e} &=& \approx \left|\frac{G^2}{\Omega^2} \int_0^\infty dk
\left( 1 + \frac{\kappa}{2\Omega A'(r_{e})} \right) e^{ -
\frac{2i\nu}{A'(r_{e})} \ln\left( \frac{\kappa}{2\Omega A'(r_{e})}
\right) }  e^{ \frac{2i\Omega}{A'(r_{e})} \left( 1 -
\frac{\kappa}{2\Omega A'(r_{e})} \right)}\right|^2, \nonumber \\
&=& \left|\frac{G^2}{\Omega^2} \int_0^\infty dk  \left( 1 +
\frac{k}{2\Omega A'(r_{e})} \right) k^{-\frac{2i\nu} {A'(r_{e})}}
e^{-ik}\right|^2, \nonumber \\
&=& \frac{4\pi G^2 \nu}{ A'(r_{e}) \Omega^2}\left(\left( 1 -
\frac{\nu}{\Omega}\right)^2 +\frac{A'^2(r_{e})}{4\Omega^2}\right)\times
\frac{1}{e^{\frac{4\pi \nu}{ A'(r_{e})} - 1}}. \label{EMP}
\end{eqnarray}
The s parameter $\Omega$ used to designate the atomic frequency in the
definition (\ref{NV}) is crucial  which will be ventilated in the following Sec. V.
Since $\Omega>> A(r_{e})$ we have the following simplified form of
the transition probability
\begin{equation}
 P_(e) \approx \frac{4\pi G^2 \nu A'(r_{e})}{e^\frac{4\pi
\nu}{f'(r_{e})} - 1}.
\end{equation}
Our goal is to observe the KR black hole's Lorentz violation effect.
Utilizing the frequency as the observer perceives it while standing
 at infinity will help us better understand it. Using the gravitational
 redraft factor, we can get this observed frequency, which we will refer to as $\nu_o$.
\begin{equation}
\nu=\frac{\nu_{o}}{\sqrt{A(r)}} \cong
\nu\left(r-r_{e}\right)^{\frac{1}{2}}. \label{RSF}
\end{equation}
We, therefore, can cast the expression of transition probability
in terms of the obsessive frequency of the emitted photon  using
the relation (\ref{RSF}):
\begin{eqnarray}
P_(e, \nu_o) &= &\frac{4\pi G^2 \nu_{o}}{ A'(r_{e}) \Omega^2}\times
\frac{1}{e^\frac{4\pi \nu_o}{ A'(r_{e})}-1} \nonumber \\
&=&\frac{4\pi G}{2\nu}  \frac{(r - r_+) A'(r_+)}{A'(r_{e})
\Omega^2} \frac{1}{e^{\frac{4\pi \nu \sqrt{(r - r_{e})
A'(r_+)}}{A'(r_{e})}} - 1}. \label{EMPF}
\end{eqnarray}
A straightforward calculation leads to the absorption probability
as well which reads
\begin{eqnarray}
P_{a} = \frac{4\pi G^2 \nu}{ A'(r_{e}) \Omega^2}\left( (1 +
\frac{\nu}{\Omega^2}) +\frac{A'^2(r_{e})}{4\Omega^2}\right)\times
\frac{1}{1-e^{\frac{4\pi \nu}{ A'(r_{e})}}}. \label{ABP}
\end{eqnarray}

\section{Computation of Modified HBAR entropy}
By connecting quantum effects close to the event horizon to thermal
entropy creation, Hawking radiation plays a key role in the deep
linkages between gravity, quantum theory, and statistical mechanics that have
been unfolded through the study of black hole thermodynamics. Beyond this thermal
image, however, acceleration-induced excitations that contribute extra entropy are
also experienced by quantum fields interacting with in-falling matter close to the
horizon. This wider entropy creation is captured by the idea of HBAR entropy,
which was first proposed in [29] and takes into account both thermal and non-thermal
quantum processes close to black hole horizons.

Motivated by string theory and other models of quantum gravity, we investigate
HBAR entropy in a Lorentz-violating Kalb-Ramond  black hole spacetime, where the
violation of Lorentz symmetry generates new physical phenomena. There may be
detectable departures from the conventional black hole thermodynamic picture
as a result of these changes to the radiation spectrum and entropy dynamics.

A system of two-level atoms that interact with the quantum fields close to the
 black hole as they fall past the event horizon is examined. We may represent
 the cumulative change in the field density using the density matrix formalism as follows:
\begin{equation}
\Delta\varrho= \sum_n\delta \varrho_n=\Delta N \delta\varrho~,
\label{DRHO}
\end{equation}
where $\Delta N$ is the number of in-falling atoms and $\delta \varrho $ the
change from a single interaction. By tracing over the atomic degrees of freedom,
we extract the HBAR entropy and analyze how Lorentz-violating terms modify the
entropy flux and radiation profile.

Our analysis highlights key deviations from conventional Hawking radiation,
offering insights into the interplay between symmetry breaking and black hole
thermodynamics, and pointing toward potential observational signatures of
Lorentz-violating effects.
In this occasion, the  rate of fall into the event horizon denoted
by  $\xi$ is define by
\begin{equation}
\frac{\Delta N}{\Delta \tau}=\xi~. \label{RFALL}
\end{equation}
We consider here a two-level atom with transition frequency $\Omega$ which is
assumed to be falling in the event Horizon
of the KR black hole. Remember that it is the same $\Omega$ which we have already
 introduced within the definition (\ref{NV})
in the preceding Sec. IV. Now the use of equations (\ref{DRHO},\ref{RFALL})
leads us to
write
\begin{equation}
\frac{\Delta \varrho}{\Delta \tau}=\xi\delta\varrho~.\label{XII}
\end{equation}
We now bring into action of the master equation for the density
matrix due to  Lindblad \cite{LIND}
\begin{equation}
\begin{split}
 \frac{d\varrho}{dt}=&-\frac{T_{A}}
{2}\left(\varrho a^\dagger a+b^\dagger a \varrho-2a\varrho a^\dagger\right)\\
&-\frac{\Gamma_{exc}}{2}\left(\varrho a a^\dagger+a a^\dagger
\rho-2a^\dagger\varrho a\right),
\end{split}
 \label{MEQ}
\end{equation}
where the rates of excitation and absorption are represented by $T_{E}$ and $T_{A}$,
 respectively. The relation
$T_{\left(E/A\right)}=\xi P_{\left(E/A\right)}$ between $T_{E}$ and $T_{A}$ is known
 to exist, with $P_{\left(E/A\right)}$provided in Equations. Both Eqns (\ref{ABP})
and (\ref{EMP}, \ref{EMPF}). The following expression is obtained
by taking the quantum average of Eqn. (\ref{MEQ}) with respect to
an arbitrary state $|n\rangle$.
\begin{equation}
\begin{split}
\dot{\varrho}_{n,n}=&-T_{A}\left(n\varrho_{n,n}-(n+1)\varrho_{n+1,n+1}\right)
\\&-T_{E}\left((n+1)\varrho_{n,n}-n\varrho_{n-1,n-1}\right)~.
\end{split} \label{AME}
\end{equation}
To obtain the precise expression of the HBAR entropy, the steady-state
 solution is helpful. In Eqn.(\ref{AME}), it enables us to
 utilize the condition $\dot{\varrho}_{n,n}=0$.  The relationship
 between $\rho_{1,1}$ and $\varrho_{0,0}$ for $n=0$ is now as follows.
\begin{equation}
\varrho_{1,1}=\frac{T_{E}}{T_{A}}\varrho_{0,0}~. \label{RDEN}
\end{equation}
If we repeat this procedure  $s$ time, we land on to
\begin{equation}
\varrho_{n,n}=\left(\frac{\Gamma_{E}}{\Gamma_{A}}\right)^s\varrho_{0,0}~.
\label{GN}
\end{equation}
Condition $Tr(\varrho)=1$ is now executed to find the explicit
expression of  $\rho_{0,0}$  in terms of $TG_{E}$ and $T_{A}$ as
follows. The condition $Tr(\varrho)=1$ leads us to write
\begin{equation}
\sum\limits_m \varrho_{m,m}=1,
\end{equation}
that provides
\begin{equation}
\varrho_{0,0}\sum_n\left(\frac{T_{E}} {T_{A}}\right)^s=1,
\end{equation}
which ultimately results in
\begin{equation}
\varrho_{0,0}= 1-\frac{T_{E}}{T_{A}}. \label{RLTION}
\end{equation}
Using $\varrho_{0,0}$ from the above  Eqn. (\ref{RLTION}),
the von-Neumann entropy for the system is computed.
\begin{equation}
\varrho^{\mathcal{S}}_{n,n}=\left(\frac{T_{E}}{T_{A}}\right)^s
\left(1-\frac{T_{E}}{T_{A}}\right).
 \label{ROS}
\end{equation}
Note that for the KR black hole
\begin{eqnarray}
\frac{T_{E}}{T_{A}} &\cong& e^\frac{-4\pi\nu}{c A'(r_{e})}\left[ 1 - 2 \frac{\nu}{\Omega} + \frac{(A'(r_e))^2}{\Omega^2} \left( \nu - \frac{1}{4} \right)\right]\nonumber \\
&\cong& e^{\frac{-8\pi G M \nu (1 - l)^2}{c^3}}\left[1 - 2 \frac{\nu}{\Omega} + \frac{c^4}{4 G^2 M^2 (1 - l)^4 \Omega^2} \left( \nu - \frac{1}{4} \right)\right],
 \label{VONN}
\end{eqnarray}
where $r_{e}=\frac{2G(1-l)\mathcal{M}}{c^2}$. To obtain Eqn.
(\ref{VONN}), we have introduced the  assumptions $\nu\ll\omega$.
 The von Neumann entropy and its rate of change for the system
are given by \cite{PAGE0}
\begin{align}
S_\varrho &=-k_B\sum\limits_{n,\nu}\varrho_{m,m}\ln\varrho_{m,m},
\\
\dot{S}_\varrho
&=-k_B\sum\limits_{m,\nu}\dot{\varrho}_{m,m}\ln\varrho_{m,m}\cong
-k_B\sum\limits_{m,\nu}\dot{\varrho}_{n,n}\ln\varrho_{m,m}
\label{VON}
\end{align}
where in case of the rate of change, we have replaced $\rho_{m,m}$
by the steady state solution $\varrho_{m,m}^{\mathcal{S}}$. Using
the form of $\varrho_{m,m}^{\mathcal{S}}$ from Eqns.(\ref{VONN},
\ref{VON}), we obtain the analytical form of $S_\varrho$ as
follows
\begin{eqnarray}
\dot{S}_\rho &\cong& -k_B\sum\limits_{m,\nu}m\dot{\rho}_{m,m}
\ln\biggr[ e^{\frac{-8\pi G M \nu (1 - l)^2}{c^3}}\left[1 - 2 \frac{\nu}{\Omega} + \frac{c^4}{4 G^2 M^2 (1 - l)^4 \Omega^2} \left( \nu - \frac{1}{4} \right)\right]\biggr]\\
 &\cong&
\biggr[\frac{8\pi G M  (1 - l)^2}{c^3} + \frac{2}{\Omega} + \frac{c^4}{4 G^2 M^2 (1 - l)^4 \Omega^2} \biggr]k_B\sum\limits_{\nu}\dot{\bar{n}}_\nu\nu\nonumber \\
 &\cong&
\left[\frac{8\pi GM\nu(1-l)^2}{c^3}+4 \frac{\nu}{\Omega} \left( 1 - \frac{c^4}{16 G^2 M^2 (1 - l)^4 \Omega^2} \right)\right]\frac{k_{B}\dot{m}_{KR}c^2}{\hbar}
 \label{SDOT}
\end{eqnarray}
where $\bar{n}_\nu$ is the photon flux generated from two-level
atoms in the vicinity of the event horizon of the black hole. In the above expression
$\frac{\nu}{\Omega^2}$ has been neglected. The
net loss of energy due to emitted photons is given by
\begin{equation}
\hbar\sum\limits_\nu \dot{\bar{n}}_\nu\nu=\dot{m}_{KR}c^2.
\end{equation}
The surface area of the black hole under consideration is
\begin{equation}
A_{KR}=4\pi r_{e}^2 = \frac{16\pi G^2 M^2 (1 - l)^2}{c^4} .
\label{AreA}
\end{equation}
The time derivative of $A_{KR}$ yields
\begin{equation}
\dot{A}_{KR}=\frac{32\pi G^2(1 -
l)^2\mathcal{M}\dot{\mathcal{M}}}{c^4}.
\end{equation}
Consequently, the rate of change of mass can be expressed as
\begin{equation}
\dot{\mathcal{M}}=\dot{m}_{KR}+\dot{m}_{a}.
 \end{equation}
Here, $M$ denotes the  mass of the black hole, and $\dot{m}_{KR}$
represents the rate at which the rest mass changes of the black hole as a
result of photon emission \cite{PAGE0}. We now use the phrase to define the
rate at which the black hole's area changes as a result of the photons it emits.
\begin{equation}
\dot{A}_{KR}=\frac{32\pi G^2(1 -
l)^2\mathcal{M}\dot{m}_{KR}}{c^4}. \label{RCA}
\end{equation}
The computation above leads us to the conclusion that, in the
absence of atoms going into the black hole, $A_{KR}$ is the black
hole's area. $A_{at}$ can therefore be regarded as vanishingly
small. The following interpretation applies to this outcome.
Before reaching the event
 horizon, the atom releases radiation. This allows for the separation of the black
 hole entropy linked to an atom and that linked to HBAR radiation from an atom.
 It is therefore acceptable that, in the absence of atoms going into the black hole,
 $A_{KR}$ equals the area of the entire black hole. Thus, the closed-form expression of $\dot{S}_\varrho$ in terms of $A_{KR}$ is as follows:
\begin{equation}
\dot{S}_\varrho =\frac{d}{dt}\biggr[F(A_{KR})\biggr]. \label{DEN}
\end{equation}

Here $F(A_{KR})$ is a function of the area of the KR black hole
which can be easily obtained from using the expression (\ref{RCA})
One of  novel findings our study is the explicit revelation of the
dependence on the Lorentz violation factor in Eqn. (\ref{DEN}).
Additionally, we note that two more $\hbar$-dependent terms appear
in Eqn (\ref{DEN}),
 which modifies the coefficient of the second term in Eqn.
 (\ref{SDOT}). According to previous publications, these terms are
 probably the result of expanding the analysis beyond the near-horizon
  approximation \cite{BAKE1, BAKE2, HAW1, HAW2, PK}.

\section{Conclusion}
The interaction between a two-level atom and a spherically
symmetry, yet Lorentz symmetry violating black hole arising from
Kalb-Ramond gravity  has been investigated in this effort. The
fact that a mirror surrounds outer edge of the black hole to keep
falling atoms from coming into contact with the Hawking radiation
allows us to analyze the entire system. Furthermore, the mirror
ensures that, from the viewpoint of an external observer, the
initial condition of the field seems to be vacuum-like. Our
findings demonstrate how effects that defy Lorentz invariance
significantly alter fundamental physics ideas and thermodynamic
properties.

Our investigation focuses on two primary aspects: analyzing the
 probability of excitation in the atom-field system and determining the rate
 of change in the Horizon Brightened Acceleration Radiation. A Planck-like
  component in the probability of excitation signifies the release of actual
  photons. Oddly, the Lorentz-violating effects brought about by the KR gravity model
  change this Planck-like factor significantly. These quantum gravity corrections are
  included in the modified frequency, suggesting a departure from the conventional
   Equivalence principle. Despite the fact that the Equivalence Principle is applicable
   in a broader context, our findings indicate a glaring violation because of the presence
   of components that defy Lorentz invariance. These results are in line with previous
    observations that studied the modifications brought about by the application of the
    Generalized Uncertainty Principle \cite{VEQUIV} and the Bumblebee Gravity Model \cite{OUR}.
    The interplay between quantum corrections for gravity and fundamental  ideas associated to
    physics is highlighted by the communication between effects that defy Lorentz invariance
    and changes brought about by the application of the Generalized Uncertainty Principle.
     As observed in our earlier work on the bumblebee effect \cite{OUR}, the violation of
     the Lorentz symmetry associated with the KR gravity model is linked to the breach of
     the equivalency principle .

The investigations suggests that it has nothing to do with conformal symmetry breaking.
 However, both the Lorentz symmetry violation factor is crucially connected to the
 Planck factor associated to the radiation that signals brings equivalency principle under
 question in the line of studies where generalized uncertainty principle was implemented to
 take quantum correction in indirect manner. This study demonstrates the important role of
 Lorentz-violating quantum gravity corrections, offering new perspectives on black hole
 physics and the fundamental ideas of equivalence in gravity theories where spontaneous
 symmetry violation causes Lorentz violation.

It should be mentioned that although the equivalency principle is a basic idea in physics,
 it can be violated when Lorentz violation enters holding the hand of  indirect quantum correction.
  Additionally, it has been observed that equivalence principle  gets threatened when the
  Generalized Uncertainty Principle is amended to take in to account the  same indirect quantum
  correction yet in different perspective. However, it is currently impossible to pinpoint exactly what is causing the equivalency principle to be broken. It does necessitate a great deal of investigation. However, one may hypothesize that it might be the outcome of quantum correction because Lorentz violation is not present in every instance where violations of equivalence principle was observed, while indirect quantum correction was there in every occurrence. Although attempts of quantization of gravity are being persued in several ways from multiple perspectives, yet the concrete, unequivocal forecast would only be made if formal quantum gravity came within our reach. This is still a long way off.

\end{document}